\documentclass[runningheads]{llncs}
\usepackage{color,soul}
\usepackage{microtype}
\usepackage[T1]{fontenc}
\usepackage{graphicx}
\usepackage[backend=biber,style=ieee]{biblatex}
\usepackage{color}

\usepackage{hyperref}

\usepackage{booktabs}
\usepackage{multirow}
\usepackage{array}

\begin{document}
\title{Towards LLM-Assisted Architecture Recovery for Real-World ROS~2 Systems: An Agent-Based Multi-Level Approach to Hierarchical Structural Architecture Reconstruction}
\titlerunning{LLM-Assisted Architecture Recovery}
%
\author{
Dominique Briechle\orcidID{0009-0000-2610-3399}\thanks{Authors are listed alphabetically; author order does not reflect individual contribution levels.} \and
Raj Chanchad\orcidID{0009-0001-1051-9303} \and
Tobias Geger\orcidID{0009-0004-4469-534X} \and
Ruidi He\orcidID{0009-0006-3849-7659} \and
Dhruv Jajadiya\orcidID{0009-0004-4142-6179} \and
Dhruv Kapadiya\orcidID{0009-0005-2759-9812} \and
Andreas Rausch\orcidID{0000-0002-6850-6409} \and
Meng Zhang\orcidID{0000-0002-9831-9356}}

\authorrunning{D. Briechle et al.}
%
\institute{Institute for Software and Systems Engineering, Clausthal University of Technology, Clausthal-Zellerfeld 38678, Germany\\
\url{https://www.isse.tu-clausthal.de/}}

\maketitle              
\begin{abstract}  
Explicit software architecture models are essential artifacts for communicating, analyzing, and evolving complex software-intensive systems. In ROS~2-based robotic systems, however, structural (de-)composition and integration semantics are often only implicitly encoded across distributed artifacts such as source code and launch files, making recovery of hierarchical architecture particularly difficult. Existing approaches mainly focus on node-level entities and communication wiring, while providing limited support for recovering hierarchical structural (de-)composition across multiple abstraction levels.

In this paper, we extend our previously proposed blueprint-guided LLM-assisted architecture recovery pipeline for ROS~2 systems through two major enhancements: (1) refined prompting to improve the consistency and controllability of architecture synthesis, and (2) a staged recovery strategy based on multi-level intermediate architectural representations that incorporate the atomic ROS node list and launch file dependencies, thereby enabling structurally constrained reconstruction across multiple abstraction levels.

The approach is evaluated on a real-world automated product disassembly system based on cooperative robotic arms and heterogeneous ROS~2 artifacts. Compared to our previous work, the considered case study exhibits substantially higher integration complexity and richer functionality. The results demonstrate improved structural consistency, scalability, and robustness of architecture recovery, while also revealing remaining challenges related to dynamic integration semantics in large-scale ROS~2 systems.

\end{abstract}
\section{Introduction}

Explicit software architecture models capture abstractions beyond implementation details and support communication, analysis, documentation, maintenance, and evolution of complex software-intensive systems. Established frameworks such as C4~\cite{brown2013software}, arc42~\cite{starke2019arc42}, and Kruchten's 4+1 View Model~\cite{469759} organize architectural descriptions across abstraction levels and viewpoints. Component- and code-oriented views, in particular, require hierarchical structural (de-)composition to represent subsystem boundaries, dependencies, and communication relationships consistently~\cite{brown2013software}.

In practice, however, architecture models are often incomplete, outdated, inconsistent, or absent~\cite{ali2018architecture,rost2014software}. Long-lived systems evolve through incremental implementation, continuous integration, and localized design decisions, causing architectural drift. This drift reduces comprehensibility, maintainability, and traceability, motivating automated architecture recovery from implementation and configuration artifacts.

While this problem is common in many software-intensive systems, it is particularly pronounced in distributed cyber-physical systems, especially ROS~2-based robotic and autonomous applications. In such systems, hierarchical composition of runtime node instances is rarely available as a complete and maintained architecture model. Instead, subsystem boundaries, node instances, communication relations, namespaces, remappings, and launch-induced compositions are implicitly encoded across source code, build files, launch files, and parameter files. Consequently, hierarchical ROS~2 architecture recovery must reconstruct structure from dispersed evidence rather than from an explicitly maintained model.

Recovering such architectures requires more than syntactic analysis: it requires interpreting integration semantics across abstraction levels, including launch-induced composition, namespace propagation, remapping, and runtime node instantiation. LLMs have therefore attracted interest for architecture recovery and documentation generation because they can infer higher-level structural and semantic relationships from heterogeneous software artifacts, e.g.,~\cite{semarch}. However, without system-specific architectural knowledge, recovery lacks guidance on the intended viewpoint, admissible ROS~2 elements, valid structural relations, and repository evidence required to justify recovered abstractions. But as a result, unconstrained LLM-based recovery may produce implausible, incomplete, inconsistent, hallucinated, or structurally unsupported elements. These limitations motivated a constrained recovery approach that makes system-specific architectural knowledge explicit and uses it to guide LLM-assisted synthesis.

In our previous work~\cite{benchat2026modelingrecoveringhierarchicalstructural}, we introduced a UML-based modeling concept for hierarchical structural ROS~2 architectures and used it as a blueprint for LLM-assisted recovery. Based on this blueprint, the recovery approach combines deterministic structural extraction with LLM-based synthesis constrained by explicit architectural contracts derived from ROS~2 structural conventions. The evaluation showed that ROS~2-specific architectural knowledge supports structurally valid multi-level reconstruction with traceability to repository artifacts. However, recovery quality decreased for repositories with complex launch-induced composition and distributed integration semantics, especially for namespace propagation, remappings, executable-to-node mappings, and hierarchical subsystem composition.

This paper retains the previously introduced UML-based modeling concept as the target blueprint, but refines the recovery process through a staged strategy with explicit intermediate artifacts. Instead of synthesizing subsystem structures directly from raw repository artifacts, the proposed approach reconstructs architectural information incrementally. It first identifies atomic ROS~2 nodes and node-local communication interfaces, then uses launch-dependency and node-instance representations to capture launch-time instantiation, namespaces, remappings, communication relations, and subsystem structures. The launch-dependency representation makes launch-file inclusion and launch-to-node containment explicit before system-level synthesis. By making intermediate recovery evidence explicit, the approach reduces semantic ambiguity, improves traceability, and increases the controllability of LLM-assisted architecture synthesis.

The proposed approach is evaluated on a real-world executable ROS~2 robotic system with richer functionality and more heterogeneous artifacts than the controlled examples in the previous study~\cite{benchat2026modelingrecoveringhierarchicalstructural}. The evaluation analyzes whether staged, blueprint-guided recovery improves structural correctness, robustness, and scalability under integration-intensive conditions, while identifying remaining limitations in recovering complex launch-induced composition.

The remainder of this paper is structured as follows. Section~2 reviews related work on ROS~2 architecture recovery and LLM-assisted software engineering. Section~3 introduces the case study and its repository characteristics. Section~4 presents the staged recovery approach, including intermediate representations, blueprint-guided synthesis, and prompt-engineering concepts. Section~5 reports the evaluation results. Section~6 concludes the paper and discusses limitations and future work.


\section{State of the Art}
By design, ROS~2 is a decentralized, component-oriented framework in which different functionalities are represented via nodes. With the pub-sub pattern at its core, these nodes are interconnected to one another by topics, services, and actions. However, the artifacts necessary to integrate a ROS~2 system consist of the different setup, launch, and CMake files, which, therefore, contain integral information necessary for understanding the system's architecture. Although already several approaches exist for modelling of ROS~2 architectures, they vary considerably in terms of their support regarding the implementation of the ROS~2 core concepts. 

One of the prominent examples includes MeROS, developed by T.Winiarski in 2023 \cite{Winiarski_2023}, which provides a SysML (Systems Modeling Language)-based metamodel for the organizations of different architectural artifacts of cognitive robot systems. Further current challenges in manual modelling efforts are pointed out, highlighting the necessity of automated architectural recovery tools \cite{Winiarski_2023}. P. Kumar et al. \cite{7416545} introduced with ROSMOD a model-driven approach for developing large-scale robotic systems. Wanninger et al. \cite{Wanninger2021} propose with ROSSi a graphical programming interface resulting in a tool for graphical development of ROS~1 projects. With ROSpec, P. Canelas developed a domain-specific language, allowing the specification and verification of ROS-based systems \cite{10.1145/3763169}. An approach for the modelling of nodes and systems via the Architecture Analysis and Design Language (AADL) is presented in \cite{7926539}. 

LLM-based systems for system architecture generation have been used in different contexts for the generation of architectural artifacts. For example, CodeBERT, using general-purpose presentations as foundations, is used for code documentation generation \cite{feng2020codebertpretrainedmodelprogramming}. CIAO (Code In Architecture Out), introduced by M. De Luca, is based on an LLM-based framework, which uses a repository as a foundation for the generation of the documentation of system-level architectural artifacts \cite{deluca2026ciaocodearchitecture}. A method for the semi-automated generation of software architecture artifacts is proposed by T. Eisenreich et al. \cite{10.11453643660.3643942}, followed by an exploratory analysis point out the current challenge of the approach to generate PlantUML outputs. 
A. Tagliaferro et al. have evaluated as well the performance of LLM-based generation of UML component diagrams with promising results \cite{11014997}. The ability to generate workflow models for different domains via LLMs was investigated by J. Xu et al. \cite{10.11453691620.3695360}.

The overall approach to using LLMs for generating architectural artifacts has, therefore, shown promising results \cite{11344283}. Although LLMs have already been used to comprehend ROS~2 systems, the recovery of architectural artifacts remains challenging \cite{duits2026largelanguagemodelsassist}. Current approaches are, however, limited in their overall ability to represent implicit hierarchical natures like the one of the ROS~2 framework. 

This leads to several limitations: Using LLMs for the recovery of architectural artifacts comes with certain risks regarding the LLM's ability to hallucinate. For an architectural ecosystem like ROS~2 with its distributed structure incorporating launch, build, and code artifacts, this increases the possibility of incorrect architectural abstractions. Hatahet et al. have therefore proposed several counter-measures, which could raise the quality of abstraction through the injection of domain-specific knowledge \cite{hatahet2025generatingsoftwarearchitecturedescription}, including enhancement through structured prompting. 

For ROS~2, the authors have investigated the applicability of these options by presenting a blueprint-based approach for ROS~2 \cite{benchat2026modelingrecoveringhierarchicalstructural}. 
The previously introduced blue print is, therefore, enhanced by two major factors to booster its performance: Through the developed stagged recovery workflow and improved prompting structure, the LLM agents are enable to produce more precise abstractions.


\section{Case Study: BrickByBrick System for Automated Dissembly}
In contrast to the case studies investigated in \cite{benchat2026modelingrecoveringhierarchicalstructural}, we aim to demonstrate the updated pipeline's capability to generate architecture artifacts of more complex, real-world ROS~2 systems. The system selected for this purpose was developed to enable an automated system to disassemble various building brick constructions in an adaptive  manner. Constructions unknown to the system are, therefore, placed into the system, which then carries out the disassembly planning as well as the necessary actions for dismantling. Consisting of three core areas for sensing, planning, and acting, the system features a variety of different hardware-based system components, which needed to be integrated into the ROS~2-based command \& control system. These hardware components include: (1) two six-degree-of-freedom (6-DOF) robot manipulators, labeled Alice and Bob, each equipped with a two-finger gripper as end effector and a controller for robot motion; (2) two 2.5D depth cameras, each mounted on one robot arm and connected to a central processing unit; and (3) two conveyor belts, each equipped with two light barriers, one at the beginning and one at the end. Each conveyor belt is connected to the controller of one robot.

\begin{figure}
    \centering
	\includegraphics[width=0.9\textwidth]{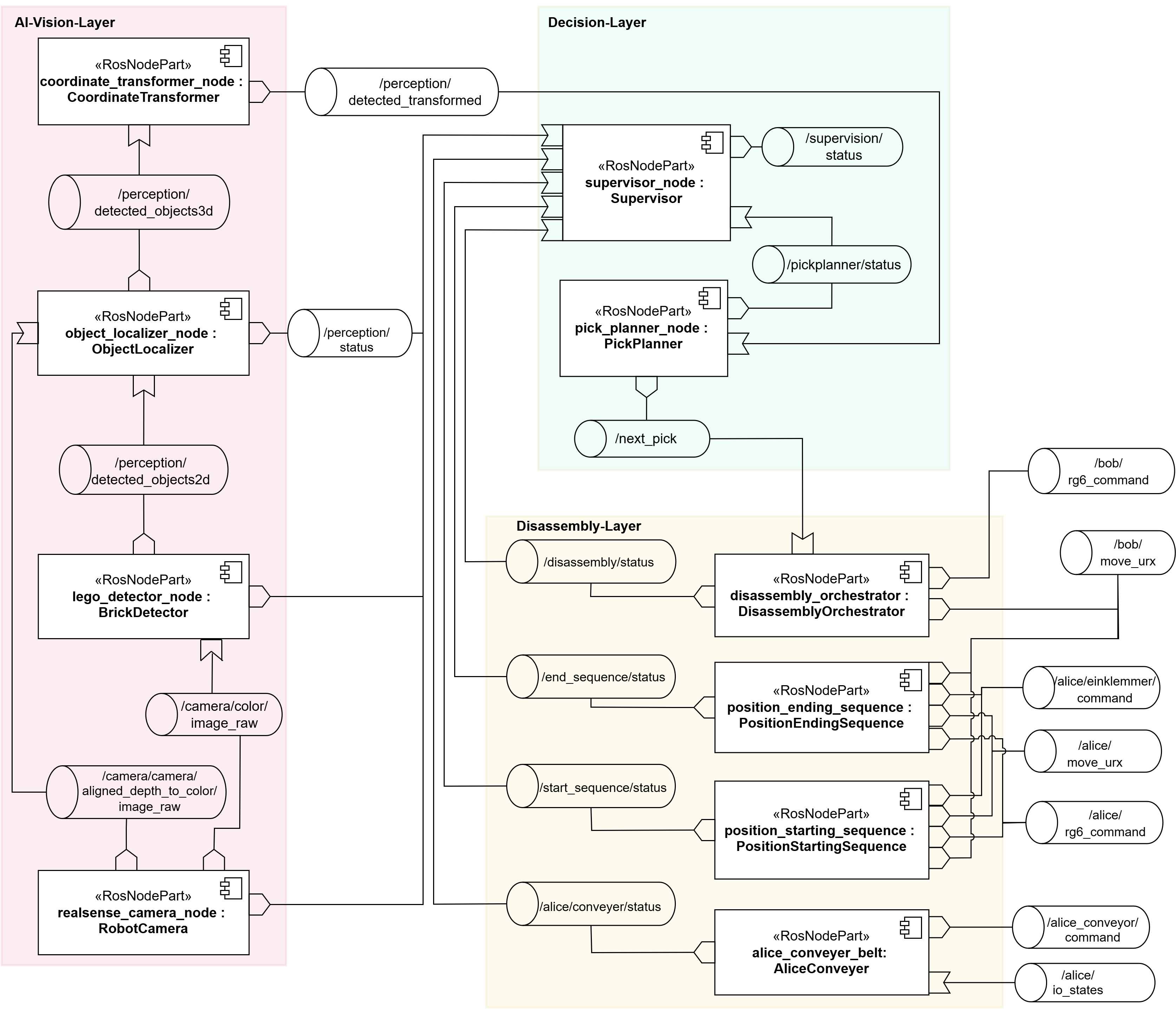}
	\caption{System architecture of the BrickByBrick disassembly system.}
	\label{fig:sysarch}       
\end{figure}

As shown in Fig. 1, the system's ROS~2-based middleware architecture is separated into three main components, responsible for the detection of the products for the disassembly and sorting task (\textit{AI-Vision-Layer}), the planning of the distinguished process (\textit{Decision-Making-Layer}), and the execution of the same via the manipulators (\textit{Disassembly-Layer}). The middleware system is connected through the \textit{Disassembly-Layer} to a subsequent \textit{Control Subsystem}, which integrates the controllers for both the robots, their corresponding end effectors, the system's sensors, and the conveyor belts \cite{10949991}. However, the subsystem is not part of the \textit{BrickByBrick Disassembly System} core architecture and is therefore not investigated in this scenario.
Initially triggered by a light barrier, the \textit{AliceConveyor} starts its hardware system to deliver the brick construction to the robot manipulator Alice.   
The \textit{RobotCamera}, as part of the \textit{AI-Vision-Layer}, takes over the responsibility to record the images of the incoming brick constructions and publishes the RGB and depth information. The subscribing \textit{BrickDetector}, which segments the bricks in the brick construction. The information regarding the detected objects is published, and the \textit{ObjectLocalizer}, subscribing to the corresponding topic, sorts the bricks depending on their hight in the construction. The adjunct \textit{CoordinateTransformer}, subscribing to the topic, transforms the coordinates with the information regarding the brick's height.

The generated information of the \textit{AI-Vision-Layer} is transmitted to the nodes of the \textit{Decision-Making-Layer}. Here the \textit{PickPlanner} contains an algorithm which decides the object to pick and sets the orientation phase for the gripper in order to avoid collisions in the picking process. 
The \textit{Decision-Making-Layer} contains as well the \textit{Supervisor}, responsible for monitoring the disassembly process by checking on the different node status. In case of a malfunction, the supervisor terminates the process and resets the system. 

The disassembly actions are executed via the \textit{Disassemby-Layer}, containing the nodes for controlling the manipulators of the hardware system as well as the conveyor belts. The core functional commands for the brick construction's disassembly are executed via the \textit{DisassemblyOrchestrator}, steering the robot and triggering the grasping process. The start and end of the disassembly process is triggered by the corresponding \textit{PositionStartingSequence} and \textit{PositionEndingSequence}, which are on their part controlled by the \textit{Supervisor}.


\section{Staged Blueprint-Guided LLM-Assisted Architecture Recovery}

\subsection{Baseline Concept: Blueprint-Guided Recovery from the Previous Study}

This subsection briefly recalls the blueprint-guided recovery concept introduced in our previous work~\cite{benchat2026modelingrecoveringhierarchicalstructural}, which serves as the baseline for the refinements presented in this paper. The underlying UML-based modeling concept is reused without modification. It defines the target architectural vocabulary for hierarchical structural ROS~2 architecture models and constrains the admissible elements and relations produced during LLM-assisted synthesis.

Following the distinction between design and integration phases proposed in~\cite{10.1145/3763169}, the blueprint separates source-level node definitions from launch-time instantiation and composition. In the design phase, atomic ROS~2 nodes are implemented in source files and define typed communication interfaces such as publishers, subscribers, service clients, and service servers. In the integration phase, launch artifacts instantiate and configure these node definitions through executable mappings, node names, namespaces, remappings, launch-file inclusions, and nested launch structures, thereby determining the effective runtime architecture and subsystem hierarchy.

The modeling concept realizes this separation by distinguishing source-level node definitions from launch-time node instances and subsystem compositions. In the considered architectural view, an atomic ROS~2 node is treated as the smallest architectural unit and is not further decomposed into internal architectural subcomponents. An \emph{AtomicRosNodeClassifier} represents the implementation-level definition of such a node together with its typed communication ports. Since ROS~2 launch files assemble and configure node instances at runtime and may also include other launch files, launch-defined composition scopes are modeled as higher-level subsystems. A \emph{ComposedRosNodeClassifier} represents such a subsystem, whose internal structure consists of \emph{RosNodePart}s. A \emph{RosNodePart} denotes a launch-time instance typed either by an \emph{AtomicRosNodeClassifier} or by another \emph{ComposedRosNodeClassifier}, thereby supporting nested subsystem hierarchies induced by launch-file inclusion. It records integration-related information such as node name, executable, namespace, and remapping configuration. Explicit namespace modeling captures hierarchical name scoping and distinguishes relative from absolute node, topic, and service names. This separation provides the architectural contract for constraining recovery and tracing recovered elements from implementation evidence to launch-time integration contexts.

In the baseline recovery process, this modeling concept was operationalized as a blueprint that constrained architecture synthesis. Deterministic static extraction first identified atomic ROS~2 node definitions from source files and produced a structured node inventory aligned with the \emph{AtomicRosNodeClassifier} concept. LLM-assisted component-level synthesis then generated PlantUML models for individual atomic nodes by mapping extracted node definitions and communication interfaces to blueprint-defined classifiers, ports, and typed communication relations. LLM-assisted system-level synthesis combined the recovered component models with launch and configuration artifacts and mapped launch-time evidence to \emph{RosNodePart} and \emph{ComposedRosNodeClassifier} structures. PlantUML served as the concrete textual representation of the recovered architecture models and as the basis for deterministic validation against the blueprint-defined structural constraints.

Our previous evaluation of the baseline concept showed that the blueprint reduced unsupported generated elements and supported structurally valid multi-level reconstruction. However, recall decreased for repositories with complex launch-induced subsystem composition, namespace propagation, remapping, and implicit integration semantics. This indicated that the main remaining challenge was not the target modeling concept itself, but the completeness and controllability of integration-phase reconstruction.

The present paper therefore keeps the UML-based modeling concept unchanged and refines only the recovery process. The key extension is the introduction of explicit intermediate artifacts that expose node, launch, and communication dependencies otherwise distributed across ROS~2 repository artifacts. In particular, the enhanced process makes source-level node identities and component-level interfaces explicit, exposes launch dependencies and runtime node instances, and records namespace scopes, remapping relations, and system-level communication relations before final system-level synthesis. These artifacts provide structured context for LLM-assisted synthesis and reduce ambiguity during system-level reconstruction. The following subsections describe how these artifacts are constructed and progressively integrated into the final architecture model.

\begin{figure*}[t]
    \centering
    \includegraphics[width=0.95\textwidth]{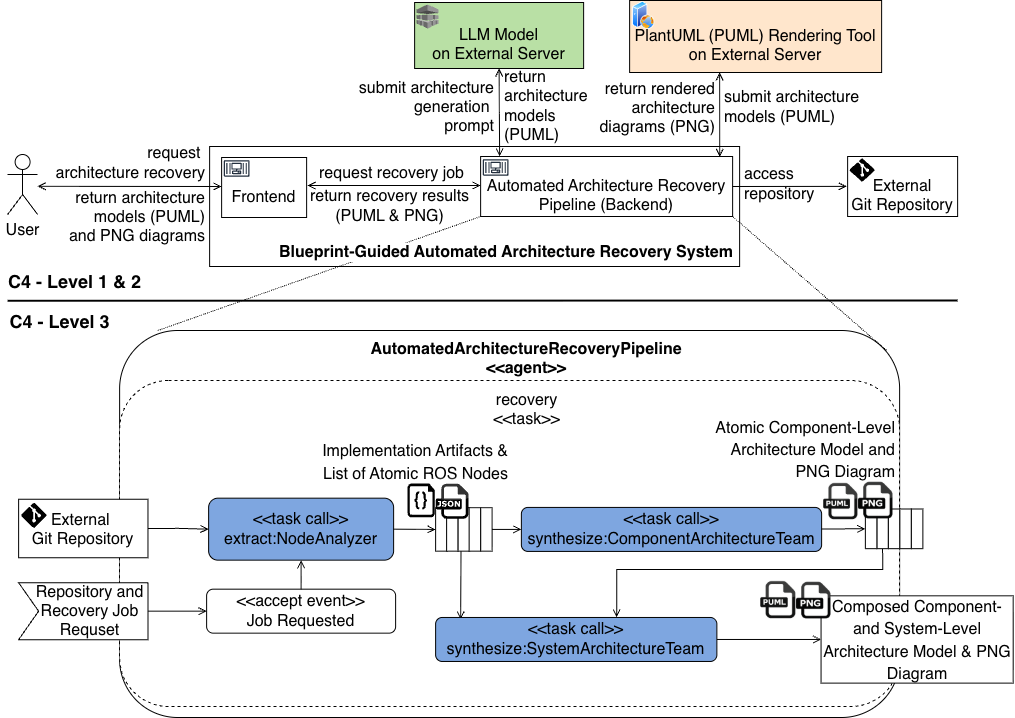}
    \caption{Blueprint-guided automated architecture recovery system and staged backend pipeline. The system receives a repository and recovery job request, uses an external LLM for architecture generation and PlantUML for rendering, and decomposes recovery into NodeAnalyzer, ComponentArchitectureTeam, and SystemArchitectureTeam stages that produce atomic and composed architecture models.}
    \label{fig:pipeline1}
\end{figure*}

\begin{figure*}[t]
    \centering
    \includegraphics[width=0.95\textwidth]{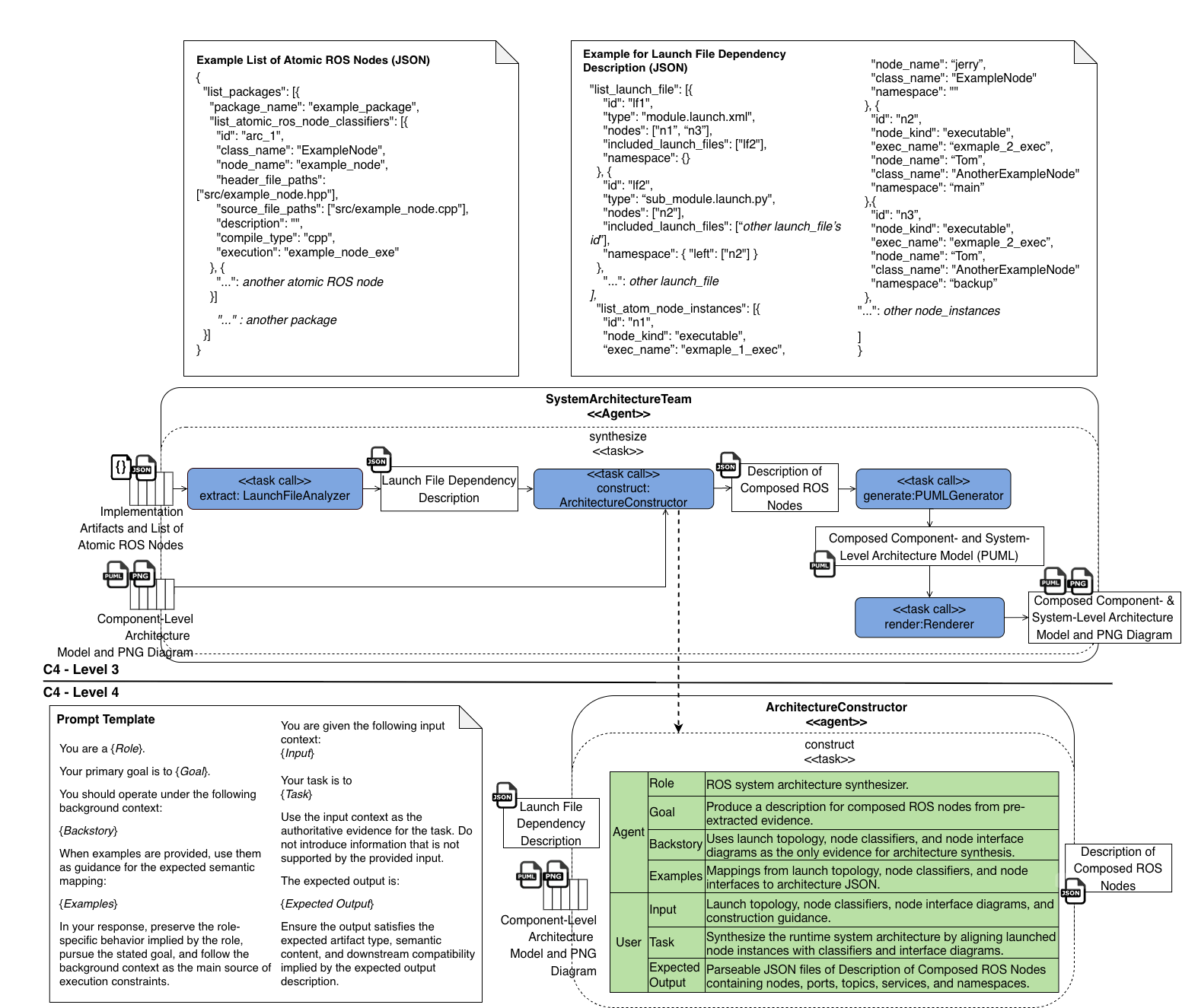}
    \caption{Refined staged recovery workflow for ROS~2 system-level architecture reconstruction. The workflow introduces two explicit JSON artifacts, the List of Atomic ROS Nodes and the Launch File Dependency Description, which provide structured prompt context for constructing and rendering the final composed architecture model.}
    \label{fig:pipeline2}
\end{figure*}

\subsection{Refined Staged Recovery Strategy and Prompt-Contract Design}

To improve semantic clarity and explainability in blueprint-guided recovery, the refined process decomposes architecture reconstruction into explicit intermediate artifacts and constrains LLM-assisted synthesis through prompt contracts. The UML-based modeling concept remains unchanged; only the recovery process is refined. The staged strategy first makes source-level atomic node definitions explicit, then materializes launch-file dependency context, and finally aligns these structured inputs with component-level interface models to synthesize the composed-component and system-level architecture diagram. The following paragraphs describe the two central intermediate artifacts and explain how they are embedded into the prompt contract for architecture synthesis.

\paragraph{List of Atomic ROS Nodes (JSON).}

The \texttt{NodeAnalyzer} agent produces a JSON-based List of Atomic ROS Nodes. This artifact organizes node definitions distributed across ROS~2 packages, source files, and header files into a unified source-level node inventory. It records node identity, source ownership, implementation language, and execution entry.

The list uses a top-level \texttt{list\_packages} array. Each package entry identifies a ROS~2 package through the \texttt{package\_name} field and lists its recovered source-level node definitions in the \texttt{list\_atomic\_ros\_node\_classifiers} field.

\begin{table}[t]
\centering
\caption{Fields and corresponding meaning of List of Atomic ROS Nodes entries.}
\begin{tabular}{p{0.30\linewidth}p{0.62\linewidth}}
\hline
\textbf{Field} & \textbf{Meaning} \\
\hline
\texttt{id} & Unique identifier of the atomic ROS node classifier. \\
\texttt{class\_name} & Source-level class identity of the ROS node. \\
\texttt{node\_name} & Recoverable node name in source file. \\
\texttt{header\_file\_paths} & Header files declaring the node class or related interfaces. \\
\texttt{source\_file\_paths} & Source files implementing the node behavior. \\
\texttt{description} & Short semantic description of the node responsibility. \\
\texttt{compile\_type} & Implementation language type, distinguishing Python nodes from C++ nodes. \\
\texttt{execution} & Executable or plugin identity used by launch files or execution entry points. \\
\hline
\end{tabular}
\end{table}

Such structure aggregates repository-level source evidence into package-scoped node classifiers. In the given example, \texttt{arc\_1} uses class name, node name, source paths, and execution entry to describe the source-level identity and runtime entry of \texttt{ExampleNode}. The list acts as the source-level index of the pipeline: \texttt{ComponentArchitectureTeam} uses the class name and source file paths to locate source context and extract component-level interfaces, while \texttt{SystemArchitectur\-eTeam} later places these classifiers into system-level runtime context.

\paragraph{Launch File Dependency Description (JSON).}

After the list of atomic ROS nodes establishes the source-level node index, the \texttt{LaunchFileAnalyzer} produces a Launch File Dependency Description before \texttt{SystemArchitectureTeam} performs system-level synthesis. This JSON artifact makes launch-time organization explicit by representing launch files, include chains, node declarations, and namespace scopes as a dependency structure.

The description contains a core array, \texttt{list\_launch\_file}, whose elements correspond to launch files included in the ROS~2 repository. Each element records the directly instantiated atomic ROS~2 node instances, included child launch files, and a \texttt{list\_atom\_node\_instances} entry describing the corresponding runtime node instances, including executable or class identity, runtime node name, and effective namespace. Stable identifiers link launch-file elements to their directly instantiated node instances and included child launch-file elements.

\begin{table}[t]
\centering
\caption{Fields and corresponding meaning of launch file entries in the Launch File Dependency Description.}
\begin{tabular}{p{0.30\linewidth}p{0.62\linewidth}}
\hline
\textbf{Field} & \textbf{Meaning} \\
\hline
\texttt{id} & Unique identifier of the launch file instance. \\
\texttt{type} & Launch file name. \\
\texttt{nodes} & Node instance IDs directly instantiated by the launch file. \\
\parbox[t]{0.28\linewidth}{%
\texttt{included\_launch\_}\\[-1pt]
\texttt{files}%
} & Launch file instance IDs directly included by the launch file. \\
\texttt{namespace} & Mapping from namespace scope to node or launch file instance IDs inside that scope. \\
\hline
\end{tabular}
\end{table}

\begin{table}[t]
\centering
\caption{Fields and corresponding meaning of node entries in the Launch File Dependency Description.}
\begin{tabular}{p{0.30\linewidth}p{0.62\linewidth}}
\hline
\textbf{Field} & \textbf{Meaning} \\
\hline
\texttt{id} & Unique identifier of the node instance. \\
\texttt{node\_kind} & Runtime implementation category marker used to help distinguish Python nodes from C++ nodes. \\
\texttt{exec\_name} & Executable name, plugin name, or class name used for classifier matching. \\
\texttt{class\_name} & Source-level class identity of the ROS node. \\
\texttt{node\_name} & Runtime node name specified in the launch file. \\
\texttt{namespace} & Effective namespace of the node instance. \\
\hline
\end{tabular}
\end{table}

This structure separates source-level definitions from system-level instances. In the given example, \texttt{n2} and \texttt{n3} both originate from \texttt{exmaple\_2\_exec} and share the runtime name \texttt{Tom}, but they are represented as distinct node instances because they belong to the \texttt{main} and \texttt{backup} namespaces. Similarly, \texttt{lf1} records the child launch file \texttt{lf2} as an include dependency and records \texttt{n1} and \texttt{n3} as directly declared runtime nodes.

The Launch File Dependency Description fixes node identities, launch dependencies, node instances, namespace scopes, and repeated instantiations before final synthesis. The constructor then aligns this runtime context with the atomic node list from Agent \emph{NodeAnalyzer} and the interface models from \emph{ComponentArchitectureTeam}, and combines namespace resolution with remapping information to derive system-level communication relations.

\paragraph{Prompt Context Integration.}

The staged artifacts are integrated into final architecture construction through a prompt contract. The SystemArchitectureConstructor prompt is organized as a reusable template filled with role, goal, backstory, examples, input, task, and expected output. These fields respectively define the analytical identity, target state, judgment boundary, mapping references, admissible evidence scope, transformation steps, and output contract. Fixed template sentences connect the fields into a continuous task specification.

The List of Atomic ROS Nodes and the Launch File Dependency Description are directly embedded in the prompt instruction context. The former specifies how runtime node instances map to source-level node definitions; the latter specifies how launch file instances, node instances, include relations, and namespace scopes are referenced. Thus, the prompt contract constrains both the synthesis procedure and the input schemas, requiring structured alignment among node identity, runtime instance, namespace, and component-level interface before the final system-level architecture model is produced.

\section{Experimental Evaluation}

\subsection {Evaluation Setup with Case Study} 

The evaluation assesses the structural correctness of component and system-level architecture models recovered by the improved staged pipeline as shown in Fig.~\ref{fig:pipeline1}. Generated PlantUML than syntactic similarity. Evaluation is carried out independently at two abstraction levels: component-level  models representing individual atomic ROS~2 nodes, and a system-level composed model capturing the runtime node composition as orchestrated across source code, launch configurations, and intermediate architectural representations.



The BrickByBrick repository comprises approximately~1,500 lines of Python source code distributed across 4~packages, 10~ROS~2 node classes, and 1~launch file\footnote{https://github.com/Tobias1998-hub/BrickByBrick}. The launch configuration applies no namespace scoping and contains no nested launch file includes; all nodes are instantiated at global scope and only once.

Compared to the repositories evaluated in our prior work~\cite{benchat2026modelingrecoveringhierarchicalstructural}, this case study is substantially more complex. The 10~node implementations cover very different functional domains: computer vision, motion planning, hardware interfacing, and process supervision, yielding semantically rich node behaviors and interaction patterns that challenge uniform extraction. Additionally, no namespace boundaries are defined in the launch files, and 20 topics are communicated over all 10 nodes. As a result, recovering subsystem structure cannot rely on syntactic cues from launch artifacts alone but requires semantic reasoning over the overall communication topology. 

The reference architecture was constructed manually through systematic inspection of all source files and launch configurations and is provided as a PlantUML model in the project repository.

\subsection {Architecture Evaluation Pipeline with Implemented Metrics}
To evaluate the staged recovery pipeline, we reuse the evaluation pipeline and metric definitions from our previous work~\cite{benchat2026modelingrecoveringhierarchicalstructural}. In brief, recovered PlantUML models are compared against expert-crafted reference models by parsing both artifacts into canonicalized sets of profile-defined architectural elements and comparing them element-wise. The evaluation therefore targets semantic conformance to the architectural profile rather than surface-level syntactic similarity.

For each metric element, shared elements are counted as true positives (TP), reference elements missing from the reconstruction as false negatives (FN), and generated elements not present in the reference as false positives (FP). Since the space of possible architectural elements is open-ended, true negatives are not defined. Precision, recall, and F1 are computed from these counts and macro-averaged across the evaluated metric elements, following the same methodology and rationale as in~\cite{benchat2026modelingrecoveringhierarchicalstructural}.

The evaluated metric elements are adopted from our previous work~\cite{benchat2026modelingrecoveringhierarchicalstructural}, where they are summarized in Table~I and derived from the trace semantics of the UML-based modeling concept. In that work, explicit trace links relate repository-level evidence to architecture-level properties, and the corresponding traceable properties are operationalized as metric elements in Table~I from our previous work. Accordingly, both reference and recovered PlantUML models are parsed into canonicalized sets of such profile-defined architectural properties, enabling deterministic element-wise comparison on architectural facts rather than diagram syntax or layout.

\begin{table}[htbp]
    \centering
    \caption{Evaluation Metrics for Case Study BrickByBrick}
    \label{tab:atomic_classifier_metrics}
    \renewcommand{\arraystretch}{1.2}
    \begin{tabular}{llccc}
        \toprule
         & \textbf{Metric Element} & \textbf{Precision} & \textbf{Recall} & \textbf{F1} \\
        \midrule
        \multirow{6}{*}{ACD} &ARC name & avg = 1.0 & avg = 1.0 & avg = 1.0 \\
       & ARC stereotype          & avg = 1.0 & avg = 1.0 & avg = 1.0 \\
       & Message Type            & avg = 1.0 & avg = 1.0 & avg = 1.0 \\
       & Callback Function Name  & avg = 1.0 & avg = 1.0 & avg = 1.0 \\
       & Service Type            & avg = 1.0 & avg = 1.0 & avg = 1.0 \\
       & Service Function Name   & avg = 1.0 & avg = 1.0 & avg = 1.0 \\
        \midrule
      \textbf{ACD} & \textbf{Average Across Elements} & \textbf{avg = 1.0} & \textbf{avg = 1.0} & \textbf{avg = 1.0} \\
      \textbf{CCD} & \textbf{Average Across Elements} & \textbf{avg = 1.0} & \textbf{avg = 0.95} & \textbf{avg = 0.98} \\
        \bottomrule
    \end{tabular}
    
    \vspace{2mm}

{\footnotesize
\raggedright
\textbf{Notes}: ARC = AtomicRosNodeClassifier; ACD = AtomicClassifierDiagram; CCD = ComposedClassifierDiagram; \textit{avg} denotes the arithmetic mean
of the six vector entries, where each entry represents the precision, recall,
and F1-score of one \textit{AtomicRosNodeClassifier} instance.\par}
 
\end{table}

    
        

\subsection {Evaluation Results}

The proposed staged recovery approach was evaluated on the BrickByBrick ROS~2 case study using the same architectural recovery metrics as in our previous work~\cite{benchat2026modelingrecoveringhierarchicalstructural}. The evaluation distinguishes between the \emph{AtomicClassifierDiagram} (ACD), which captures implementation-level \emph{AtomicRosNodeClassifier} recovery, and the \emph{ComposedClassifierDiagram} (CCD), which captures launch-induced subsystem composition recovery.

At the atomic level, all evaluated architectural elements achieved average precision, recall, and F1-scores of 1.0. The evaluated ACD elements include classifier names, stereotypes, message types, callback function names, service types, and service function names. These results indicate that the refined process preserves the strong implementation-level recovery performance already observed in the earlier prototype~\cite{benchat2026modelingrecoveringhierarchicalstructural}.

At the composed level, the improved staged recovery pipeline achieved a precision of 1.0, a recall of 0.95, and an F1-score of 0.98 for the recovered CCD model. The excellent precision indicates that the generated system-level elements are structurally supported by repository evidence, while the remaining recall gap reflects a small number of missing launch-induced composition relations.

Compared with the earlier prototype results, the CCD-level performance suggests improved recovery of subsystem-level ROS~2 architecture semantics, particularly with respect to launch-induced hierarchy construction, namespace propagation, and inter-node composition recovery. Overall, the results support the feasibility of recovering high-fidelity architectural representations directly from heterogeneous ROS~2 repository artifacts.

\section{Conclusion and Outlook}

This paper presented an enhanced blueprint-guided approach for recovering hierarchical structural architecture models of ROS~2-based systems. Building on the UML-based modeling concept introduced in our previous work, the proposed approach refines the recovery process through a staged strategy with explicit intermediate artifacts. In particular, a source-level list of atomic ROS~2 nodes and a launch-file dependency description make node identities, launch inclusions, runtime node instances, and namespace scopes available before final composed-system synthesis. These artifacts provide structured context for LLM-assisted recovery and improve the controllability and traceability of architecture reconstruction.

The evaluation on the BrickByBrick case study shows that the staged process preserves strong implementation-level recovery performance, with all ACD-level metric elements achieving precision, recall, and F1-scores of 1.0. At the CCD level, the recovered composed architecture achieved a precision of 1.0, a recall of 0.95, and an F1-score of 0.98. These results indicate that the refined process can recover high-fidelity architectural representations from heterogeneous ROS~2 repository artifacts and improves subsystem-level reconstruction compared with the earlier prototype.

The remaining recall gap at the composed level shows that complete reconstruction of integration semantics remains challenging, especially when subsystem boundaries and composition relations are only implicitly encoded through launch files, communication topology, naming conventions, or project-specific integration patterns. Future work will therefore focus on formalizing the modeling concept into a more explicit UML profile or metamodel, extending recovery support for complex launch constructs, namespace propagation, remapping, and repeated node instantiation, and incorporating behavioral or runtime evidence to complement static repository-based recovery. Establishing shared ROS~2 architecture recovery benchmarks with reference models and common metrics would further improve reproducibility and cross-approach comparability.

\begin{credits}
\subsubsection{\ackname}

This work has been developed in the project “GESAL” (Research Grant Number ZW 7-87011681) and is funded by the European Regional Development Fund (ERDF) and the State of Lower Saxony. The authors would like to thank Mohamed Benchat, Mitbhai Chauhan, Meet Chavda, Nidhiben Kaswala, and Daniel Osterholz for their contributions to the original blueprint-guided ROS~2 architecture recovery study, which provided the foundation for this work. Further, the authors also would like to thank the BrickByBrick team, Mika Czekay, Erich Feifert, Konrad Günster, Marcel Lindwedel, Wladimir Miller, Lena Schartow, and Maximilian Zarna, for making the system available for evaluation.

\subsubsection{\discintname}
The authors have no competing interests to declare that are relevant to the content of this article.

\end{credits}
\printbibliography
\end{document}